\documentclass[aps,,nofootinbib,amsfonts,groupedaddress]{revtex4}
\usepackage{amsmath}
\usepackage{epsfig,epsf}
\usepackage{graphicx}
\usepackage{verbatim}
\usepackage{epstopdf}
\usepackage{bm}  
\usepackage{epsfig,epsf}
\def\beq{\begin{equation}}
\def\eeq{\end{equation}}
\def\bea{\begin{eqnarray}}
\def\eea{\end{eqnarray}}
\def\eq#1{{Eq.~(\ref{#1})}}
\def\fig#1{{Fig.~\ref{#1}}}

\newcommand{\Lb}{\left(}
\newcommand{\Rb}{\right)}
\parskip=\itemsep               

\newcommand{\nn}{\nonumber}

\newcommand{\pom}{I\!\!P}

\def\pom{{I\!\!P}}

\begin{document}

\preprint{TAUP 2962/13}

\title{Diffraction Production in a Soft Interaction Model: 
Mass Distributions}

\author{E.~ Gotsman$^{1}$,~ E.~ Levin$^{1,2}$ ~and~ U.~Maor${}^{1}$}

\affiliation{
${}^{1}$~Department of Particle Physics, School of Physics and Astronomy,
Raymond and Beverly Sackler
Faculty of Exact Science, Tel Aviv University, Tel Aviv, 69978, Israel\\
${}^2$~Departamento de F\'\i sica, Universidad T\'ecnica Federico Santa 
Mar\'\i a, Avda. Espa\~na 1680\\ and Centro 
Cientifico-Tecnol$\acute{o}$gico de Valpara\'\i so,Casilla 110-V, 
Valpara\'\i so, Chile}%

\date{\today}

\begin{abstract}
In the framework of our model (GLM) for soft interaction 
with $\alpha'_{\pom}(0)=0$, we propose a procedure based on Gribov's partonic 
interpretation of the Pomeron, which enables one to calculate the 
diffractive mass distributions in hadron-hadron scattering. Using the 
analogy with deep-inelastic scattering, we associate the Pomeron-quark 
interaction  with the Good-Walker sector of the hadron-hadron scattering, 
and the Pomeron-gluon sector with the t-channel Pomeron interactions. We 
present predicted mass distributions for the LHC energies.
\end{abstract}
\pacs{13.85.-t, 13.85.Hd, 11.55.-m, 11.55.Bq}
\keywords{Soft Pomeron, BFKL Pomeron, Diffractive Cross Sections, N=4 SYM}

\maketitle

\section{Introduction}
Recently we proposed a model for soft interactions at high energies 
that  provides a good description of  the LHC data on total, 
elastic and diffractive cross
sections\cite{GLM0,GLM1,GLM2}, as well as inclusive hadron 
production\cite{GLMINCL}. This model incorporates the main features of
 theoretical approaches to high energy interactions: viz. 
perturbative QCD\cite{QCDHE,BK,JIMWLK,KL} (pQCD)   
and N=4 SYM\cite{LIKO,BST,HIM,COCO,BEPI,LMKS}. 
The resulting features are:\\ 
i) A large value of the Pomeron intercept 
($\Delta_\pom \approx 0.2 - 0.3$) and diminishing 
$\alpha'_\pom = 0$ (pQCD and N=4 SYM);\\ 
ii) A large contribution of Good-Walker\cite{GW}(GW) mechanism to 
diffraction production (N=4 SYM);\\ 
iii) Significant triple Pomeron (3\pom) interaction 
(matching with pQCD).
\par
In Table 1 we show our predictions for the different components  
of single diffraction production. $\sigma_{sd}^{GW}$   
corresponding to the GW mechanism,  
while $\sigma_{sd}^{\pom}$ corresponds to the 
contribution of multi-Pomeron interactions to diffraction production. 
This table demonstrates that most of 
the diffractive cross section in our model stems 
from the GW mechanism,  this in accordance with the N=4 SYM.
\begin{table}
\small
\begin{tabular}{|l|l|l|l|}
\hline
~&~&~&~\\
W = $\sqrt{s}$  \,\, TeV
& $\sigma^{GW}_{sd}$ (mb)
&  $\sigma^{\pom}_{sd}$(mb)
& $\sigma_{sd} = \sigma^{GW}_{sd} +\sigma^{\pom}_{sd}$ (mb) 
\\
~&~&~&~\\
\hline
0.9     
&  8.44 
& 0.06
& 8.5
  \\
\hline
2.76 
& 9.68
&1.65
&11.33
  \\
\hline
7
&10.7 
&4.18 
& 14.88 
 \\
\hline
8
&10.9 
&4.3
&15.2
   \\
\hline
13
&  11.4
&5.6
& 17
  \\
\hline \hline 
\end{tabular}
\caption{Predictions of our model for single diffractive production 
cross sections at different energies $W$.
\label{t}}
\end{table}
A shortcoming of our approach is that 
we are unable to calculate  distributions of the produced diffractive 
mass.  This deficiency should be corrected, 
in view of the recent experimental activity at LHC, where missing mass 
distributions are planned to be measured in the near future\cite{COM}. 
The main goal of this paper is to  suggest an approach 
which will allow us to calculate
these mass distributions, based on new physical ideas.
\par 
Gribov partonic interpretation of the Pomeron\cite{Gribov} implies 
that the typical transverse momentum in the parton
cascade that describes the Pomeron can be specified in a simple parton
model by $q^2_{\perp} \approx \,1/\alpha'_\pom$.
Consequently, we believe that 
$\alpha'_\pom \simeq 0$ reflects the fact that the soft Pomeron is, 
actually, rather hard.
 The Donnachie-Landshoff Pomeron\cite{DL} has 
$\alpha'_\pom\, = \,0.25 GeV^{-2}$ leading to a scale of hardness 
of approximately $4 \,GeV^2$.
\par
Our key idea is to view a soft interacting Pomeron 
as a hard probe that measures 
the quark and gluon contents of the hadron target,  
in our case a proton (see \fig{pomprob}). 
We develop this idea, so as to be able 
to predict the mass distribution in diffraction production,
with an additional assumption that the diffractive 
GW sector is initiated by Pomeron interactions 
with quarks within the hadron, while non GW diffraction 
stems from Pomeron interactions with the hadronic target gluons.
\par
Note
 that for $\Delta_\pom \simeq 0.3$ both 
mechanisms for diffraction production, i.e. GW and non GW, 
lead to the production of diffractive mass whose values does not depend on 
the total energy\cite{GUF}. 
Our suggested approach to the diffractive mass distribution 
recovers the widely used classification, in which the 
GW mechanism is responsible for diffraction in the region of 
relatively low mass, while non GW mainly describes the 
production of high diffractive masses. 
\par
In the next section we  present simple formulae that transcribe the 
above ideas to the diffractive mass distributions. 
We aim to predict 
these mass distributions in the LHC kinematic region.  In section III,
we determine the scale of hardness for the Pomeron ($\widetilde{Q}$, 
 by comparing with the Tevatron data.  
 We find  the best value to be equal $\widetilde{Q}^2\,=
\,2\,GeV^2$.
 In the conclusions we  summarize our main results.
\section{Pomeron as a hard probe of the hadron content}
\begin{figure}[ht]
\begin{center}
       \includegraphics[width=12cm] {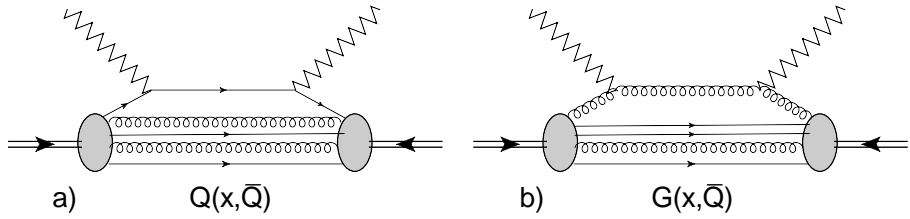}
\end{center}
\caption{The soft Pomeron as a hard probe:
\fig{pomprob}-a shows the interaction with quarks
while the interaction with gluon is depicted in \fig{pomprob}-b.
The zigzag line denotes the Pomeron. The solid and helix lines
show the quarks and gluons.}
\label{pomprob}
\end{figure}

We assume that the soft Pomeron can be viewed as a hard probe with a  
scale of hardness $\bar{Q}$. This means that the Pomeron interacts 
with quarks and gluons as a  composite (see \fig{pomprob}-a and 
\fig{pomprob}-b), 
in a way similar to that in which a virtual photon interacts in deep 
inelastic scattering processes.
Using this analogy we can write: 
\beq \label{MEQ}
\frac{d \sigma_{sd}}{d \ln\Lb M^2/M^2_0\Rb}\,\,=\,\,\sigma^{GW}_{sd}\,
q\Lb \frac{\bar{Q}^2}{ M^2 +\bar{Q}^2 }, \bar{Q}^2 \Rb/I_q\Lb 
M_{max}\Rb\,\,
+\,\ \sigma^{\pom}_{sd}\,g\Lb \frac{\bar{Q}^2}
{M^2 + \bar{Q}^2}, \bar{Q}^2\Rb/I_g\Lb M_{max}\Rb.
\eeq
In \eq{MEQ} $q\Lb x\Rb$ and $g\Lb x \Rb$ are the quark and the gluon 
distribution at the scale of hardness $\bar{Q}^2$. 
$I_q$ and $I_g$ are defined as
\beq \label{QGDEF}
I_q\,\,=\,\,\int^{M^2_{max}}_{M^2_{min}}\,\frac{d M^2}{M^2} 
q\Lb \frac{\bar{Q}^2}{M^2 + \bar{Q}^2}, 
\bar{Q}^2\Rb~~~~~~~\mbox{and}~~~~~
I_g\,\,=\,\,\int^{M^2_{max}}_{M^2_{min}}\,\frac{d M^2}{M^2} 
g\Lb \frac{\bar{Q}^2}{ M^2 + \bar{Q}^2},\bar{Q}^2\Rb.
\eeq
$M_{max}$ and $M_{min}$ are the maximal (minimal) mass that have 
been reached experimentally.

The energy variable (Bjorken $x$) for Pomeron-hadron scattering 
is equal to 
\bea \label{X}
&&0  \,=\, (p_\pom + p_{p})^2\, =\, -\, \bar{Q}^2 \,
+ \,x \,2\, p_\pom \cdot p_{h}; ~~~~~~~   p^2_\pom\, 
=\, -\, \bar{Q}^2; \nn\\
&& (p_\pom + p_{h})^2\,=\,-\, \bar{Q}^2\,
+\,   2 \,p_\pom \cdot p_{h};~~~~~~~~~~~~~~x\,
=\,\frac{\bar{Q}^2}{ M^2 + \bar{Q}^2}.
\eea
$p_\pom$, $p_{h}$ and $p_{p}$ are the momenta of the 
Pomeron, the hadron and the parton (quark or gluon) with which the Pomeron 
interacts (see \fig{pomprob}).
\begin{figure}[ht]
 \includegraphics[height=3.5cm,width=9cm] {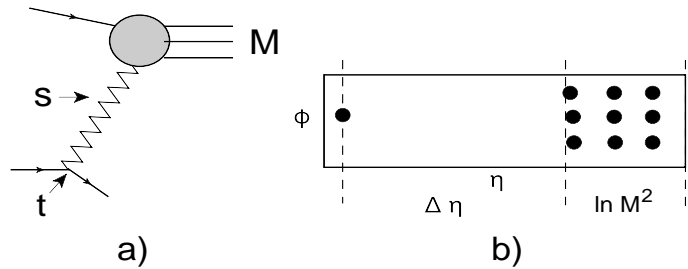}\
\caption{Single diffractive production and related kinematic variables.}
\label{diffpic}
\end{figure}

The value of $M_{min}$ can be as small as 
$M_{min} = m_{p}\,+\,m_{\pi}$.  
The value of $M_{max}$ is bounded by the condition 
that we have a Pomeron exchange. As such, the value of 
$\Delta \eta$ that corresponds to Pomeron exchange (see \fig{diffpic}) 
should be large enough, so  that one can neglect the possible 
exchanges of the 
secondary Reggeons. 
In our initial analysis
we took $\Delta \eta \,\geq \,2$.  Our model \cite{GLM0},
with $\Delta \eta_{min}\,=\,2$, suggests that the contribution
of the secondary Reggeons is approximately 50\%.
Note that, the variable $\xi$ which is usually introduced to 
describe diffraction production, is equal to
$\xi \,= \,1 - x_L = M^2/s = \exp\Lb - \Delta \eta\Rb$.
The choice of $\Delta \eta > 2$ implies that $\xi \,<\,0.05$.
At the LHC energies we took $M_{min}=1.1\,GeV$ and $M_{max}=200\,GeV$, 
which corresponds to $\Delta \eta_{min} = 7$.  For this rapidity 
the 
contribution of the secondary Reggeons amounts to  less than 10\%.
\par
\begin{figure}[ht]
\begin{tabular}{c c c}
  \includegraphics[ width=7cm] {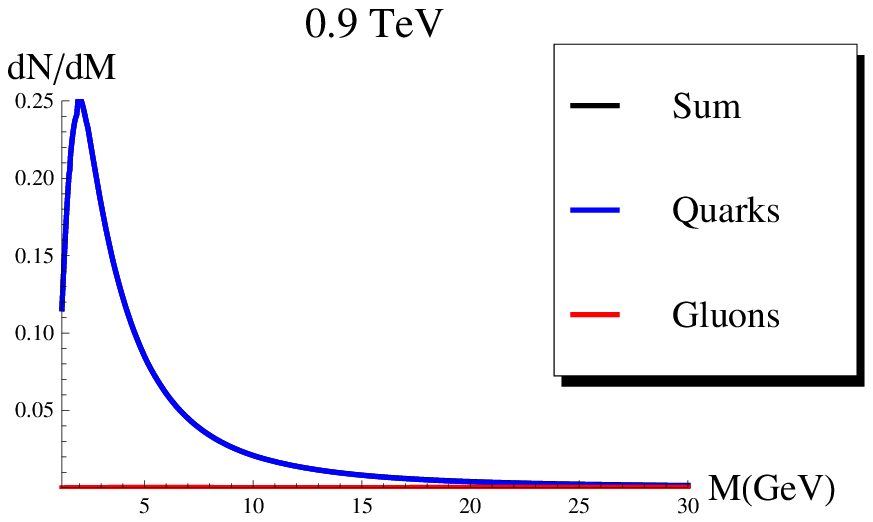} & ~~& \includegraphics[ width=7cm] {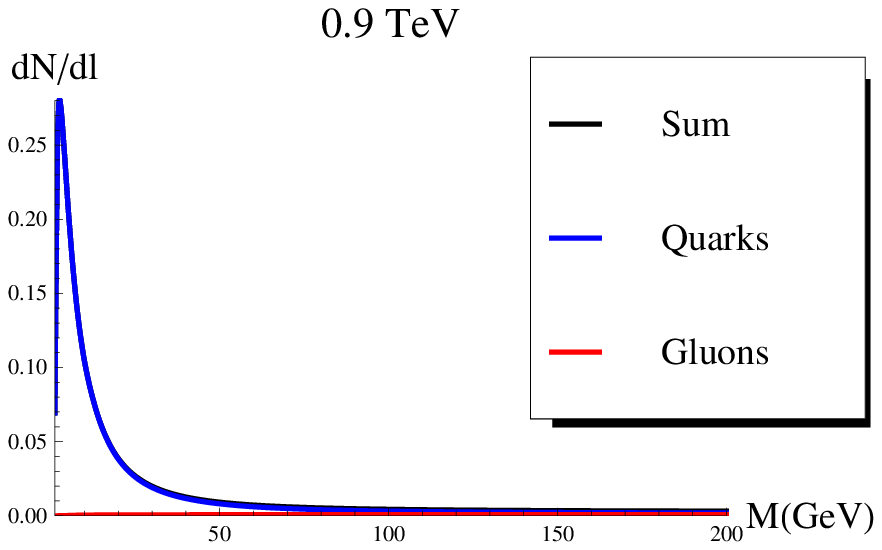}\\
    \includegraphics[ width=7cm] {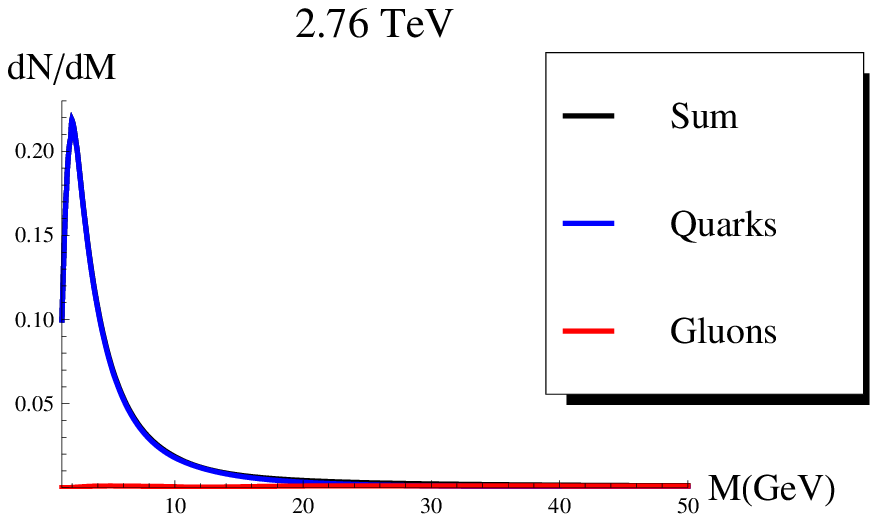} &~~&  \includegraphics[ width=7cm] {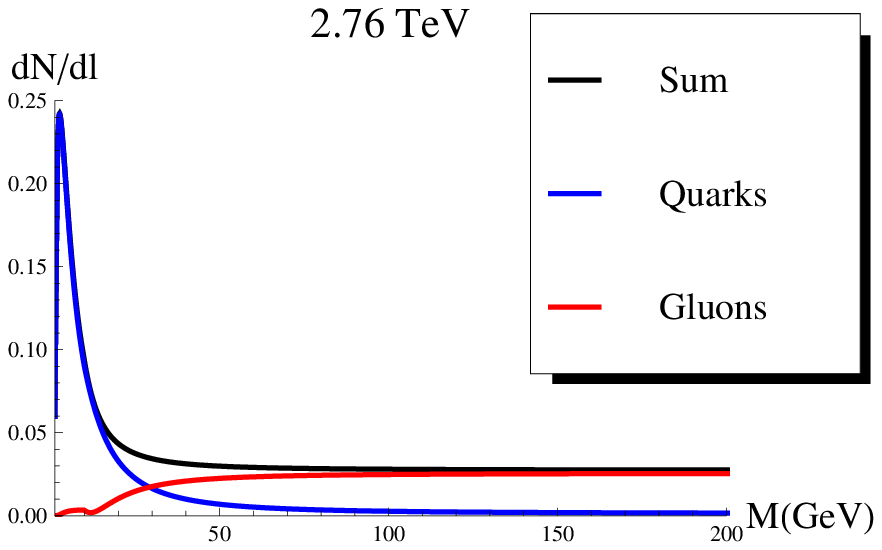}\\ 
    \end{tabular}
\caption{$d N /d M$ and $d N/d l$ with $l = \ln\Lb M^2/M^2_{min}\Rb$ 
versus $M$ for energies $W = 0.9 \,TeV$ and $2.76\,TeV$. For quark and gluon structure functions the H1-Zeus combined fit (HERAPDF01)\cite{HERAPDF}  is used. The scale of hardness for the Pomeron is taken $\widetilde{Q} \,=\,1.42\,GeV$.}
 \label{dsdm1}
\end{figure}
\begin{figure}[ht]
\begin{tabular}{c c c}
  \includegraphics[ width=7cm] {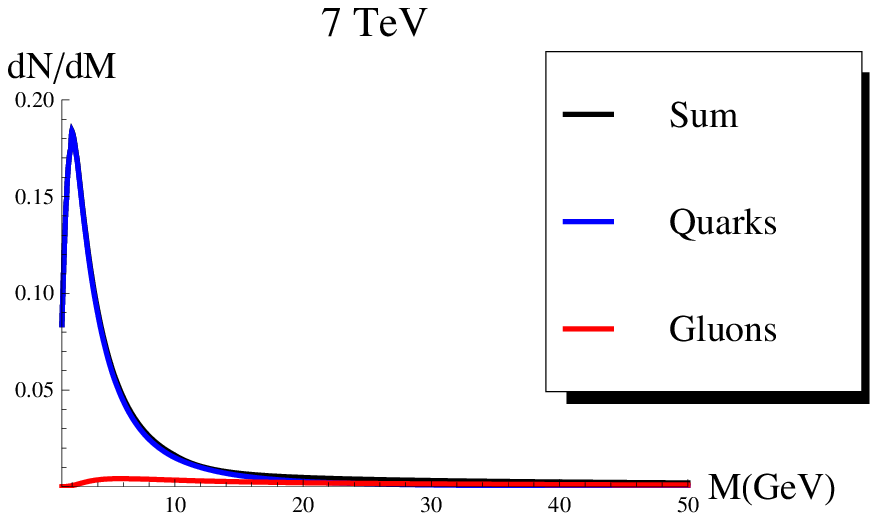} & ~~& \includegraphics[ width=7cm] {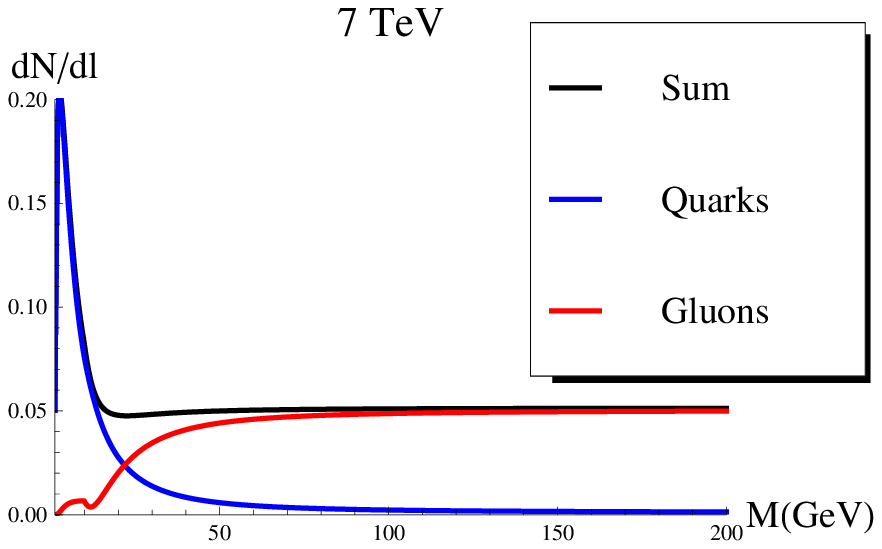}\\
    \includegraphics[ width=7cm] {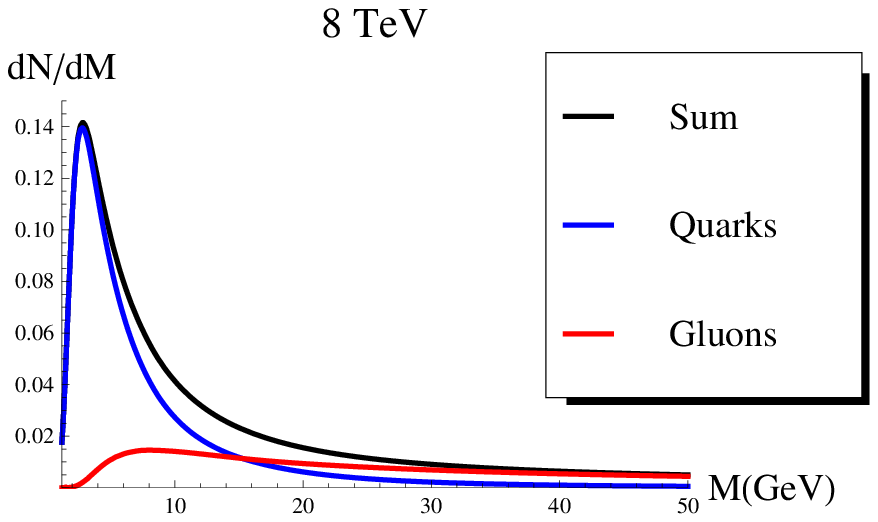} & ~~& \includegraphics[ width=7cm] {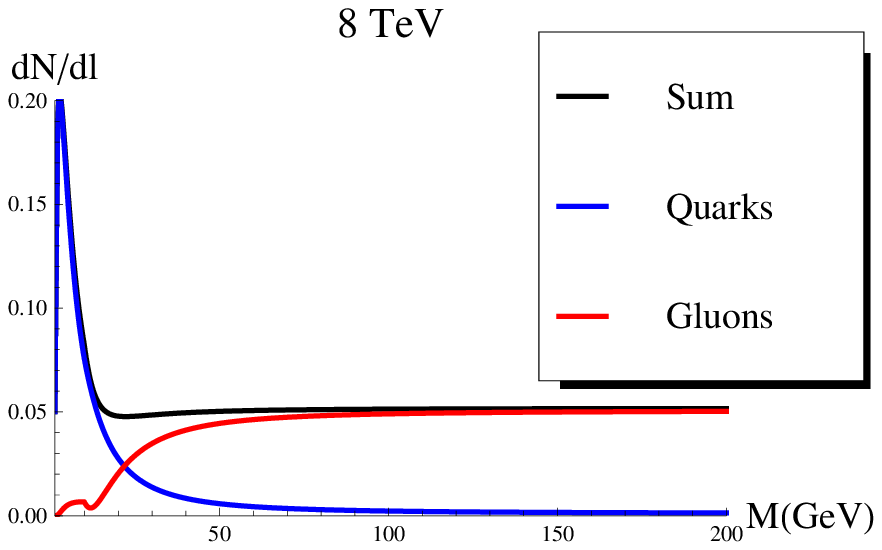}\\
    \end{tabular}
\caption{$d N /d M$ and $d N/d l$ with $l = \ln\Lb M^2/M^2_{min}\Rb$ 
versus $M$ for energies $W = 7 \,TeV$ and $8\,TeV$. For quark and gluon structure functions the H1-Zeus combined fit (HERAPDF01)\cite{HERAPDF}  is used. The scale of hardness for the Pomeron is taken $\widetilde{Q} \,=\,1.42\,GeV$.}
 \label{dsdm2}
 \end{figure}
\begin{figure}[ht]
\begin{tabular}{c c c}
  \includegraphics[ width=7cm] {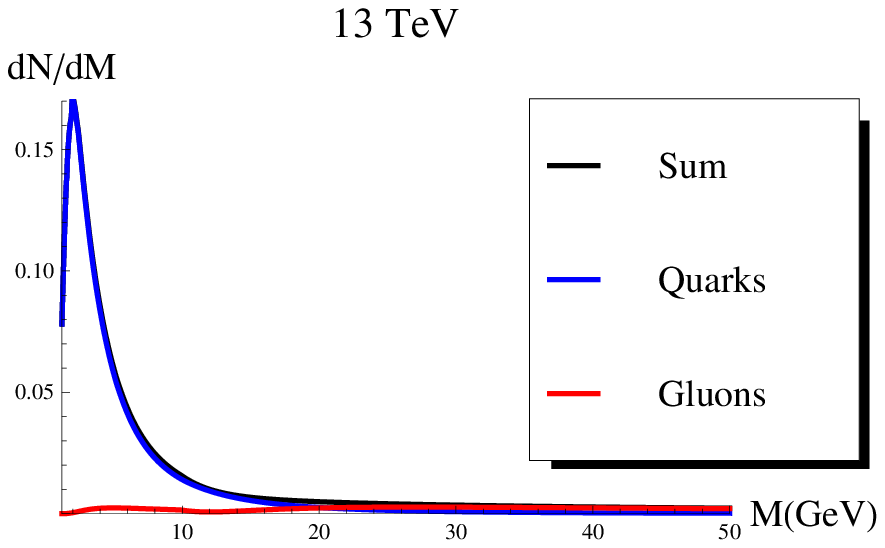} & ~~& \includegraphics[ width=7cm] {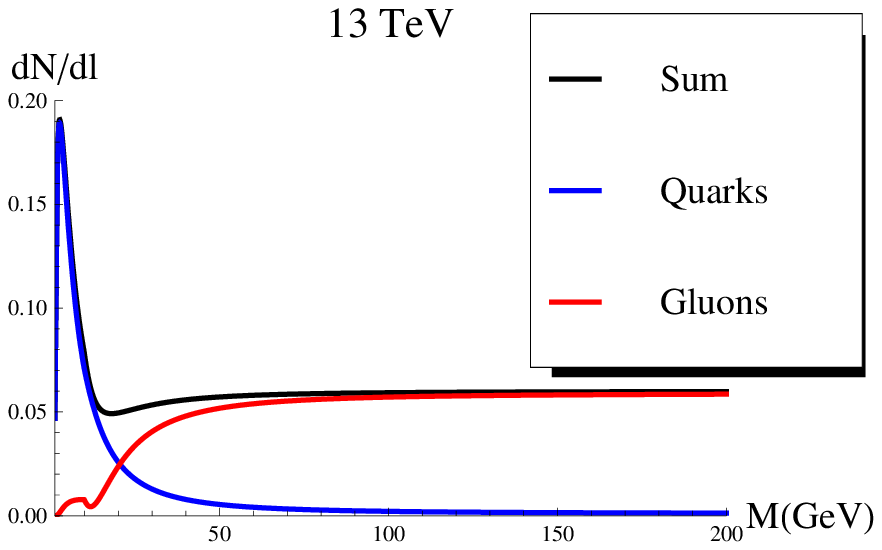}\\
        \end{tabular}
\caption{$d N /d M$ and $d N/d l$ with $l = \ln\Lb M^2/M^2_{min}\Rb$ versus  $M$ at $W = 13\,TeV$. For quark and gluon structure functions the H1-Zeus combined fit (HERAPDF01)\cite{HERAPDF}  is used. The scale of hardness for the Pomeron is taken $\widetilde{Q} \,=\,1.42\,GeV$.} \label{dsdm3}
\end{figure}

In \fig{dsdm1}- \fig{dsdm3} we plot the 
predictions\footnote{That numerical tables of our predicted mass 
distributions at LHC energies are obtainable from E. Levin.} for 
$M_{max} = 200\,GeV$ and $M_{min} = 1.1\,GeV$. 
For $\bar{Q}^2$ we choose  the value of $ 2 \,GeV^2$ from the description of th
e CDF data\cite{CDF} at the Tevatron (see the next section). 

\par
We wish to  emphasize that the  relative contribution of the 
quarks and gluons depends entirely on our model for soft interactions. 
However, the prediction turns out to be sensitive to both the value 
of the Pomeron's scale of hardness, and to the uncertainties in the 
gluon densities.
\par
\begin{figure}[ht]
\begin{tabular}{c c c}
  \includegraphics[ width=5.5cm] {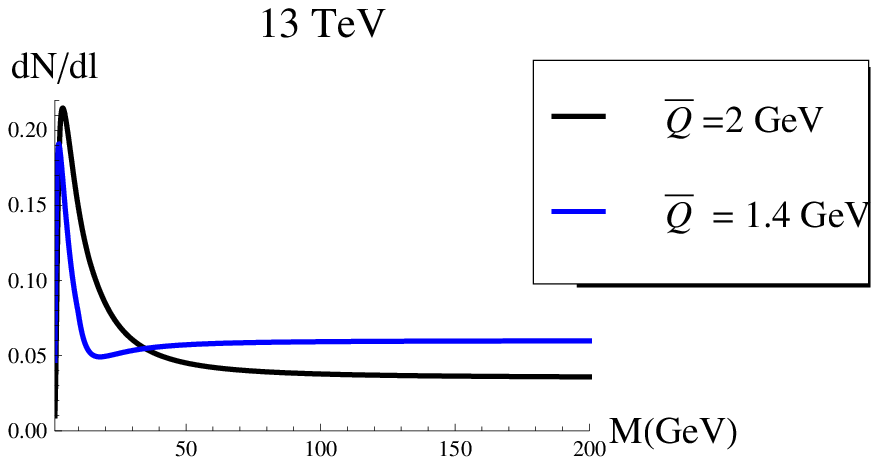} &  \includegraphics[ width=5.0cm] {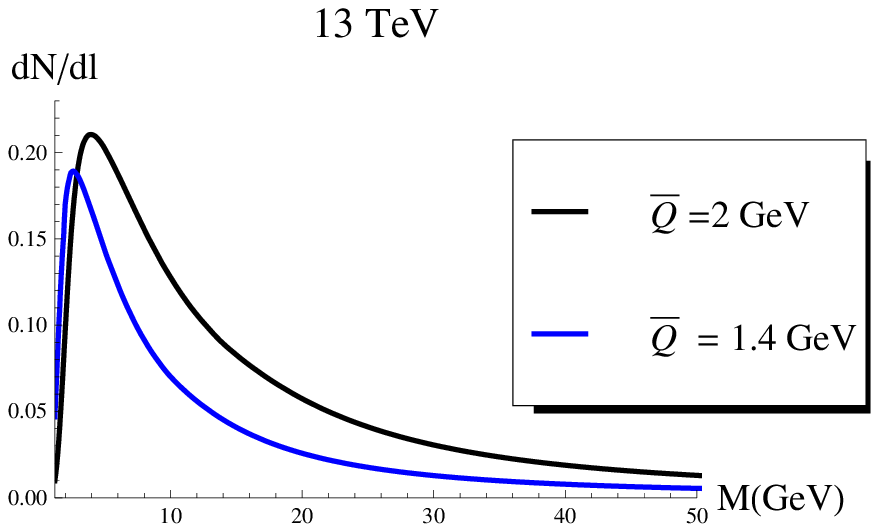}&  \includegraphics[ width=5.5cm] {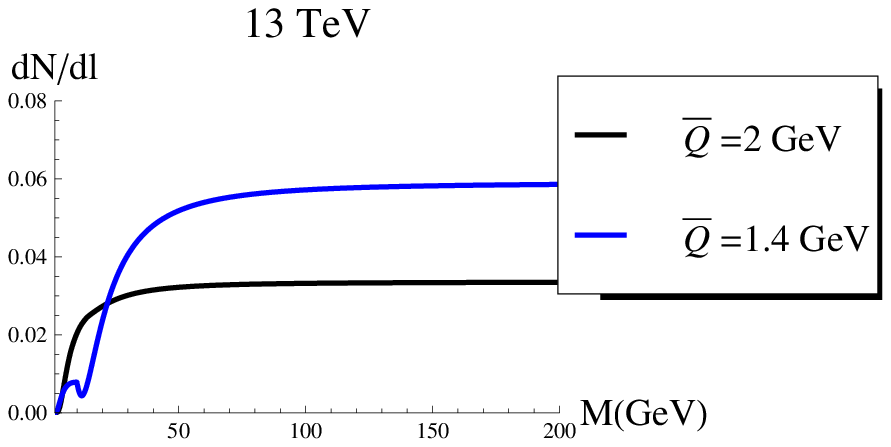} \\
        \end{tabular}
        \caption{ $d N/d l$ with $l = \ln\Lb M^2/M^2_{min}\Rb$ versus 
$M$  at $W = 13 \,TeV$ for different scales of hardness for Pomeron 
($\bar{Q}$) .}
 \label{soh}
\end{figure}
\begin{figure}[ht]
\begin{tabular}{c}
  \includegraphics[ width=7cm] {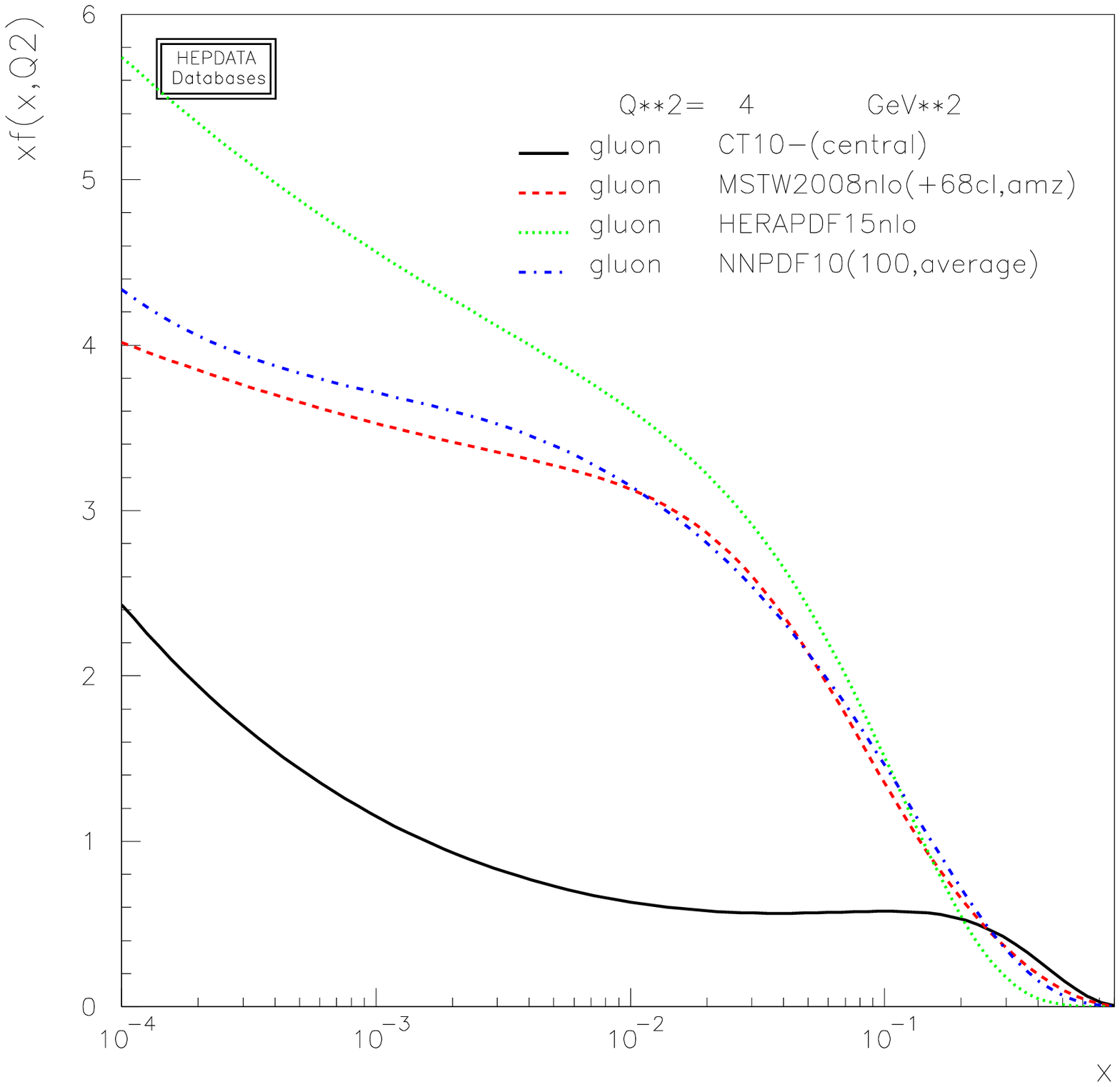}
  \end{tabular}
\caption{Gluon densities for different parameterizations. 
The  figure is taken from Durham HEP data 
(http://durpdg.dur.ac.uk/HEPDATA/).}
\label{gluons}
\end{figure}
\begin{figure}[ht]
\begin{tabular}{c c c}
  \includegraphics[ width=5cm] {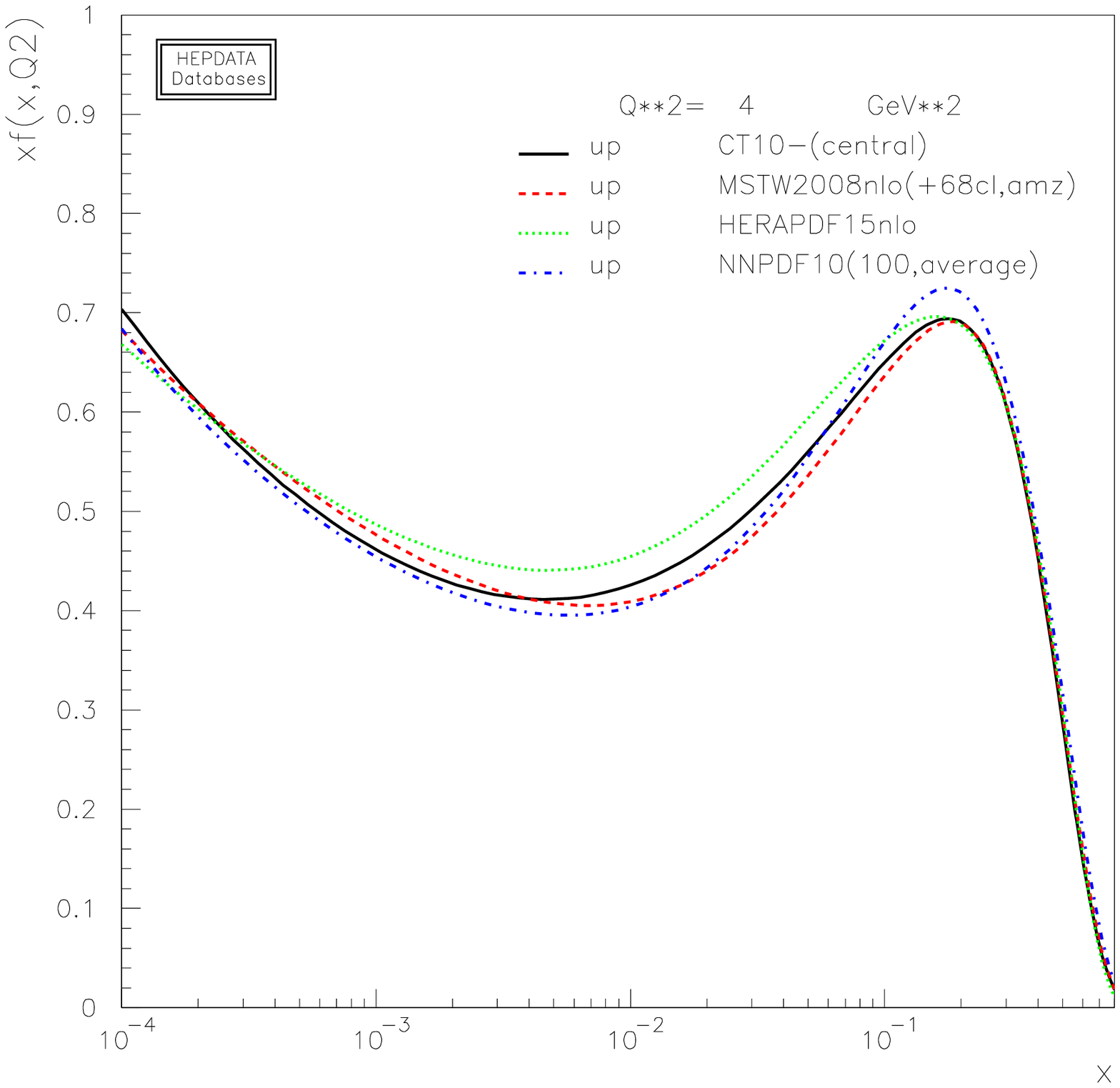} & ~~& \includegraphics[ width=5cm] {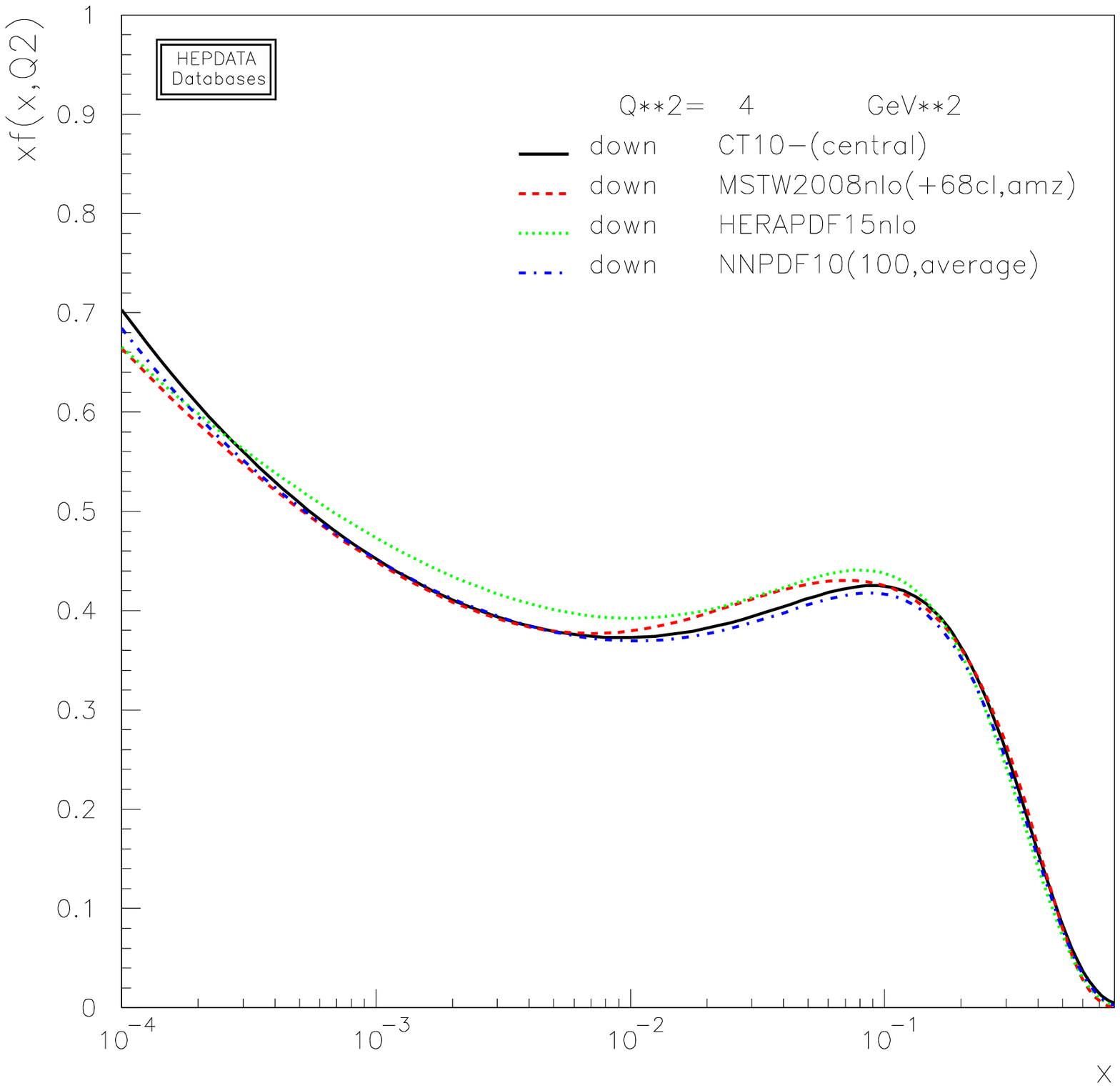}\\
\end{tabular}
\caption{ Quark densities for different parameterizations.
The figure is taken from Durham HEP data 
(http://durpdg.dur.ac.uk/HEPDATA/) .}
\label{quarks}\
\end{figure}

The resulting mass distribution depends on the Pomeron 
scale of hardness (see \fig{soh}), where we plotted the prediction for 
$\bar{Q}^2 = 4 \,GeV^2$ and $\bar{Q}^2 = 2\,GeV^2$. 
Note that, the quark  contribution is less sensitive to 
the value of the Pomeron scale of hardness, than to  
the gluon contribution, which depends crucially  on $\bar{Q}$.
There are large uncertainties in the gluon structure functions,
since these  have been extracted from the experimental data which
are only indirectly connected to the gluon densities(see \fig{gluons}).
We believe that by measuring $d N/d \ln\Lb M^2/M^2_0\Rb$, one will 
obtain additional information on the gluon densities  
which should reduce these uncertainties. 
On the other hand, the 
low mass distributions that depend on the quark densities, do not 
suffer from such uncertainties, and can be predicted rather accurately. 
(see \fig{quarks}).
\section{Comparison with the experimental data}
\begin{figure}
\begin{tabular}{c}
  \includegraphics[ width=10cm] {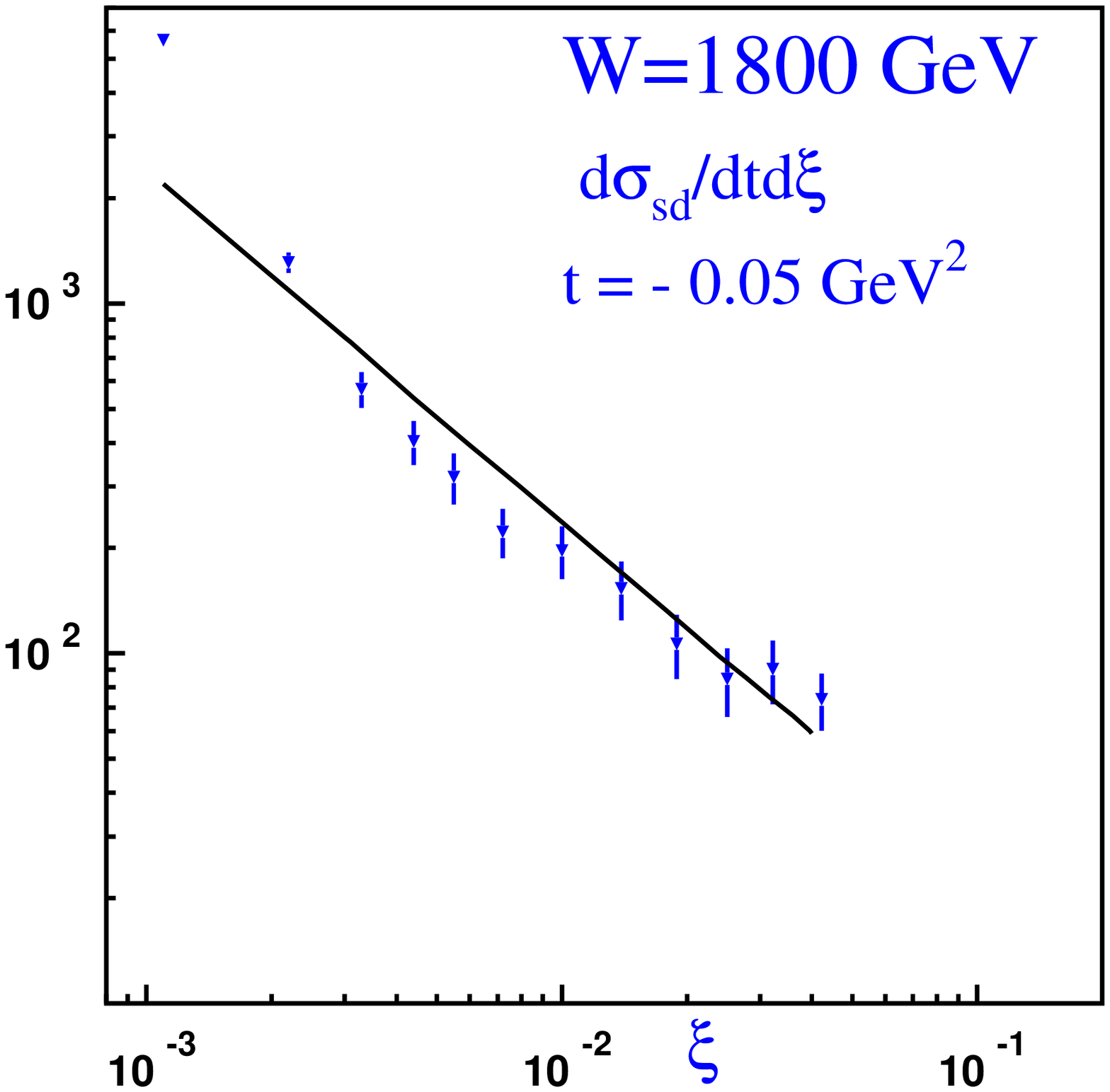}
  \end{tabular}
\caption{The comparison with the CDF data \cite{GOMO,CDF} at the Tevatron. 
$\xi = M^2/s$. The scale of hardness for the Pomeron is taken to be 
equal to $\widetilde{Q}^2\,=\,2\,GeV^2$.}
\label{tev}
\end{figure}
The mass distribution for single diffractive production has  only 
been 
measured at low energies, however, we  compare with the data 
obtained at the Tevatron\cite{GOMO,CDF}, and found even this data base 
not adequate to determine the Pomeron contribution. Even though, one can 
produce at the Tevatron
large diffractive masses, up to $M\, =\, 360 GeV$, 
we find that the contribution of the secondary
Reggeons is significant, about 50\% given $\Delta \eta_{min}\,=\,2$.
We conclude that taking  $\Delta \eta_{min}\,=\,2$  is not sufficient 
to induce 
a strong enough suppression of the secondary Reggeons contribution 
to single diffraction. 
We note that, diffractive $M= 400\,GeV$ produced 
at the LHC corresponds to $\Delta \eta = 6$. 
Therefore, we will have to wait for the diffractive 
mass distribution at the LHC to discuss
the Pomeron induced diffractive production. 
In \fig{tev} we plot the CDF data and our estimates. 
One can see from this comparison that the data support 
a scale of hardness for the Pomeron 
$\widetilde{Q}^2 \,=\,2\,GeV^2$.  Bearing in mind all uncertainties 
that stem 
from the gluon structure function and the contribution of the 
secondary Reggeons, which could change the behavior at large $\xi$, 
we consider that our attempt to describe the data is rather successful. 
\par
It is important to mention that for this comparison we  require
more input from our model. 
To this end, we use the value of the single diffraction slope 
$B^{GW}_{sd}=6.36 
\,GeV^{-2}$(see Ref.\cite{GLM0}. This slope has been calculated 
for the single diffractive GW sector. For the non GW diffractive 
Pomeron sector, 
the slope is  the same as $B^{GW}_{sd}$. For the description of the CDF data we
 use the following formula
\bea\label{MEQ1}
\frac{d \sigma_{sd}}{d \xi \, d t}\,\,&=&\,\,\frac{s}{M^2} \Big\{ B^{GW}_{sd}\exp\Lb B^{GW}_{sd}\,t\Rb \sigma^{GW}_{sd}\,
q\Lb \frac{\bar{Q}^2}{ M^2 +\bar{Q}^2 }, \bar{Q}^2 \Rb/I_q\Lb 
M_{max}\Rb\,\,\nn\\
&+&\,\, \Big( B^{\pom}_{sd}\exp\Lb B^{\pom}_{sd}\,t\Rb\,  \sigma^{\pom}_{sd}\,g\Lb \frac{\bar{Q}^2}
{M^2 + \bar{Q}^2}, \bar{Q}^2\Rb/I_g\Lb M_{max}\Rb\Big\}
\eea

\section{Conclusions}
This paper was triggered by the absence of a theoretical procedure 
to calculate the single diffractive mass distribution. 
Such a procedure is an essential ingredient in the forthcoming 
experimental analysis of the mass distribution in the diffractive 
channels, foremost the leading single diffraction channel 
To rectify this deficiency,
we have suggested that the Pomeron that has been adopted
in Refs.\cite{GLM0,GLM1,GLM2}, stems from processes with 
sufficiently large transverse momenta, and can be viewed as a hard 
probe of the constituents of the hadron. 
The simple formula of \eq{MEQ} indicates how we can use the mass 
distribution of single diffraction production, to measure the quarks 
and gluons in a hadron.
We trust that these will be helpful in understanding the soon to be available 
experimental diffractive mass data from the LHC.
\par
We  believe that the description of
$dN_{sd}/d \ln \Lb M^2/M^2_0\Rb$ 
in terms of quarks and gluons, will determine 
the Pomeron scale of hardness, as well as additional 
information on the gluon densities.

\section*{Acknowledgement}
We thank  Risto Orava, Martin Poghosyan and  Jean-Pierre Charles Revol  for dis
cussion on the experimental situation in diffractive production. This  research
 was supported by  the  Fondecyt (Chile) grants 1100648.



\begin{thebibliography}{99}
\bibitem{GLM0}
E.~Gotsman, E.~Levin and U.~Maor,
  Phys.\ Lett.\ {\bf B716} (2012) 425
  [arXiv:1208.0898 [hep-ph]];
  Phys.\ Rev.\ {\bf D85} (2012) 094007
  [arXiv:1203.2419 [hep-ph]].
\bibitem{GLM1}
 E.~Gotsman, E.~Levin and U.~Maor,
  Eur.\ Phys.\ J.\ {\bf C71} (2011) 1553
  [arXiv:1010.5323 [hep-ph]].
%
\bibitem{GLM2}
 E.~Gotsman, E.~Levin, U.~Maor and J.~S.~Miller,
  Eur.\ Phys.\ J.\ {\bf C57} (2008) 689
  [arXiv:0805.2799 [hep-ph]].
\bibitem{GLMINCL}
   E.~Gotsman, E.~Levin and U.~Maor,
  Phys.\ Rev.\ {\bf D84} (2011) 051502
  [arXiv:1103.4509 [hep-ph]];\,\,\,
  Phys.\ Rev.\ {\bf D81} (2010) 051501
  [arXiv:1001.5157 [hep-ph]].
\bibitem{QCDHE}
 F.~E.~Low,
  Phys.\ Rev.\ {\bf D12} (1975) 163;\,\,\,
 S.~Nussinov,
  Phys.\ Rev.\ Lett.\  {\bf 34} (1975) 1286;\,\,\,
E. A. Kuraev, L. N. Lipatov, and F. S. Fadin,  Sov. Phys.
JETP {\bf 45}, 199 (1977); \,\,\,
Ya. Ya. Balitsky and L. N. Lipatov,
Sov. J. Nucl. Phys.\,{\bf 28}, 822 (1978);\,\,
A.~H.~Mueller, Nucl.\ Phys. {\bf B415}, 373 (1994);
 {\bf B437}, 107 (1995);\,\,\,L. V. Gribov, E. M. Levin and M. G. Ryskin, 
Phys. Rep.\,{\bf 100}, 1 (1983);\,\,\,
A. H. Mueller and J. Qiu,   Nucl. Phys.\,{\bf B268} 427 (1986) ;\,\,\,
L. McLerran and R. Venugopalan, Phys. Rev.\,{\bf D49}, 2233, 3352 (1994);
{\bf D50},2225 (1994); {\bf D53},458 (1996); {\bf D59},094007 (1999);\,\,
L. N. Lipatov,
 Phys.\ Rept.  {\bf 286}, 131 (1997)
[arXiv:hep-ph/9610276];\,\,Sov. Phys. JETP {\bf 63}, 904 (1986) and
references therein.\,\,
\bibitem{BK}
I. Balitsky, Nucl. Phys. {\bf B463} (1996) 99; Y. Kovchegov, Phys. Rev. 
{\bf D60} (1999) 034008.
\bibitem{JIMWLK}
~J.~Jalilian-Marian, A.~Kovner, A.~Leonidov and H.~Weigert,
Phys.\ Rev.\,  {\bf D59} (1999) 014014
[hep-ph/9706377];\,\, Nucl.\ Phys.\,{\bf B504} (1997) 415
[hep-ph/9701284];
E. Iancu, A. Leonidov and L.D. McLerran,
Phys.\ Lett.\, {\bf B510} (2001) 133
[hep-ph/0102009];\,\, Nucl.\ Phys.\,{\bf A692} (2001) 583
[hep-ph/0011241];
H. Weigert,
Nucl.\ Phys.\,{\bf A703} (2002) 823 [hep-ph/0004044].
\bibitem{KL}
Yu.Kovchegov and E. Levin, 
{\it ``Quantum Choromodynamics at High Energies''}, Cambridge University 
Press, 2012 and references therein.
\bibitem{LIKO}
A.V. Kotikov, L.N. Lipatov, A.I. Onishchenko and V. N. Velizhanin,
  Phys.\ Lett.\ {\bf B595} (2004) 521
   [Erratum-ibid.\ {\bf B632} (2006) 754]
  [hep-th/0404092]  and references therein.
\bibitem{BST} 
 R.C. Brower, J. Polchinski, M.J. Strassler and C.I. Tan,
  JHEP {\bf 0712} (2007) 005
  [arXiv:hep-th/0603115];\,\,
R. C. Brower, M. J. Strassler and C. I. Tan,
{\it "On The Pomeron at Large 't Hooft Coupling"}, arXiv:0710.4378 [hep-th].
 \bibitem{HIM}
Y. Hatta, E. Iancu and A.H. Mueller,
  JHEP {\bf 0801} (2008) 026
  [arXiv:0710.2148 [hep-th]].
 \bibitem{COCO}
 L. Cornalba and M.S. Costa,
 Phys. Rev. {\bf D78}, (2008) 09010,
  arXiv:0804.1562 [hep-ph];\,\,\,
  L. Cornalba, M.S. Costa and J. Penedones,
  JHEP {\bf 0806} (2008) 048
  [arXiv:0801.3002 [hep-th]];\,\,
  JHEP {\bf 0709} (2007) 037
  [arXiv:0707.0120 [hep-th]].
\bibitem{BEPI}
B. Pire, C. Roiesnel, L. Szymanowski and S. Wallon,
  Phys.\ Lett.\ {\bf B670}, 84 (2008)
  [arXiv:0805.4346 [hep-ph]].
\bibitem{LMKS}
E. Levin, J. Miller, B.Z. Kopeliovich and I. Schmidt,
JHEP {\bf 0902} (2009) 048;\,\,
  arXiv:0811.3586 [hep-ph].
\bibitem{GW}
M.L. Good and W.D. Walker,
Phys. Rev. {\bf 120} (1960) 1857. 
\bibitem{COM}
Private communication with Jean-Pierre Charles Revol of the ALICE 
collaboration and Risto Orava of the TOTEM collaboration.
\bibitem{Gribov}
 V.N. Gribov,
  {\it ``Space-time description of hadron interactions at high energies,''}
  arXiv:hep-ph/0006158;\,\,
  Sov.\ J.\ Nucl.\ Phys.\ {\bf 9} (1969) 369 [ Yad.\ Fiz.\ {\bf 9} (1969) 640].
\bibitem{DL}
 A. Donnachie and P.V. Landshoff,
 Nucl. Phys. {\bf B231} (1984) 189; Phys. Lett. {\bf B296} (1992) 227; 
Zeit. Phys. {\bf C61} (1994) 139.
\bibitem{GUF}
G.~Gustafson,
  Phys.\ Lett.\ B {\bf 718} (2013) 1054
  [arXiv:1206.1733 [hep-ph]].
  \bibitem{CDF}
 F.~Abe {\it et al.}  [CDF Collaboration],
  Phys.\ Rev.\ {\bf D50} (1994) 5535.
  \bibitem{HERAPDF}
F.~D.~Aaron {\it et al.}  [H1 and ZEUS Collaboration],
  JHEP {\bf 1001} (2010) 109
  [arXiv:0911.0884 [hep-ex]].
  
\bibitem{GOMO}
 K.~A.~Goulianos and J.~Montanha,
  Phys.\ Rev.\ {\bf D59} (1999) 114017
  [hep-ph/9805496].
\end{thebibliography}
\end{document}